\documentclass[conference]{IEEEtran}
\usepackage{color}

\ifCLASSINFOpdf
\else
\fi
\usepackage{array}
\usepackage{caption}
\usepackage{subcaption}
\usepackage{url}

\usepackage{multirow}
\usepackage{graphicx}
\usepackage[table]{xcolor}
\usepackage{hyperref}
\usepackage[T1]{fontenc} 
\usepackage{comment}

\usepackage{cite}

\newcommand\subash[1]{\textcolor{green}{Subash : #1}}
\newcommand\ivan[1]{\textcolor{orange}{Ivan : #1}}
\newcommand\milan[1]{\textcolor{blue}{Milan : #1}}

\usepackage{amsmath,amssymb,amsfonts}
\usepackage{graphicx}
\usepackage{textcomp}
\usepackage{algorithm}
\usepackage{algpseudocode}
\usepackage{color}
\usepackage{verbatim}
\usepackage{cite}
\usepackage{textcomp}
\usepackage[table]{xcolor}
\usepackage{enumitem}
\usepackage{multicol}
\usepackage{multirow}
\usepackage{gensymb}
\usepackage{todonotes}
\usepackage{hyperref}
\usepackage{tikz}
\usepackage{tikz,lipsum,lmodern}
\usepackage[most]{tcolorbox}

\usepackage{array}
\renewcommand{\arraystretch}{1.1}
\newcolumntype{C}{>{\centering\arraybackslash}p{1em}}
\usepackage{booktabs}
\usepackage{multirow}
\usepackage{makecell}
\usepackage{caption}
\usepackage{xurl}

\usepackage{textcomp}

\newtcolorbox{boxEnv}{
center,
left=0 mm,
top = 0.25 mm,
right = 0mm,
bottom =0.25 mm,
colframe=gray!90!black,
colback=black!5!white, 
boxrule=0.5pt,
}

\def\BibTeX{{\rm B\kern-.05em{\sc i\kern-.025em b}\kern-.08em
    T\kern-.1667em\lower.7ex\hbox{E}\kern-.125emX}}
\begin{document}

\title{Impacts and Risk of Generative AI Technology on Cyber Defense}




\author{\IEEEauthorblockN{Subash Neupane, Ivan A. Fernandez, Sudip Mittal, Shahram Rahimi} Dept. of Computer Science \&  Engineering, Mississippi State University, Starkville, MS\\Email: \{sn922, iaf28\}@msstate.edu, \{mittal, rahimi\}@cse.msstate.edu\\
}

\maketitle

\begin{abstract}



Generative Artificial Intelligence (GenAI) has emerged as a powerful technology capable of autonomously producing highly realistic content in various domains, such as text, images, audio, and videos. With its potential for positive applications in creative arts, content generation, virtual assistants, and data synthesis, GenAI has garnered significant attention and adoption. However, the increasing adoption of GenAI raises concerns about its potential misuse for crafting convincing phishing emails, generating disinformation through deepfake videos, and spreading misinformation via authentic-looking social media posts, posing a new set of challenges and risks in the realm of cybersecurity. 

To combat the threats posed by GenAI, we propose leveraging the Cyber Kill Chain (CKC) to understand the lifecycle of cyberattacks, as a foundational model for cyber defense. This paper aims to provide a comprehensive analysis of the risk areas introduced by the offensive use of GenAI techniques in each phase of the CKC framework. 
We also analyze the strategies employed by threat actors and examine their utilization throughout different phases of the CKC, highlighting the implications for cyber defense. Additionally, we propose GenAI-enabled defense strategies that are both attack-aware and adaptive. These strategies encompass various techniques such as detection, deception, and adversarial training, among others, aiming to effectively mitigate the risks posed by GenAI-induced cyber threats. 


\end{abstract}


\IEEEpeerreviewmaketitle

\section{Introduction}

\IEEEPARstart{G}{}enerative Artificial Intelligence (GenAI) has become a significant and influential force in the realm of technology, empowering algorithms and models to autonomously create highly realistic content across multiple domains, including text, images, audio, and videos. This technology holds tremendous potential for positive applications across various domains, including creative arts, content generation, virtual assistants, and data synthesis, exemplified by notable innovations such as ChatGPT \cite{chatgpt}. GenAI enables content generation by automating the creation of diverse and engaging textual and visual content catering to specific styles, tones, or target audiences. Virtual assistants, on the other hand, enhance human-computer interaction by providing personalized recommendations, answering queries, and performing various tasks based on user input. Furthermore, GenAI aids in data synthesis by analyzing large datasets and generating synthetic data that closely resembles the original, facilitating machine learning algorithm training and testing while preserving privacy.

This technology, despite its immense capacity for beneficial utilization (for example, Google Bard \cite{gbrad}, Bing Chat\cite{bingchat}, and ChatGPT \cite{chatgpt}), has also introduced a new dimension of risk to the cybersecurity landscape \cite{bergman2022guiding}. As GenAI models become more pervasive, there is an increasing risk that they may be used to craft convincing phishing emails, generate realistic deepfake videos for disinformation campaigns \cite{pbsnews}, or create seemingly authentic social media posts to spread misinformation \cite{marcus2023skeptical}. According to a report published by Darktrace \cite{darktrace} in April 2023, there was a 135\% increase in the number of spam emails that featured ``\emph{markedly improved English-language grammar and syntax}'' between the months of January and February. This surge was attributed to threat actors leveraging GenAI for their phishing campaigns. Similarly, a deepfake video \cite{bbcdeepfake} depicting Ukrainian President Volodymyr Zelenskyy surfaced on online platforms, calling for his people to surrender to Russian forces in March 2023. 
This incident serves as a compelling demonstration of how deepfakes can be employed to propagate disinformation and manipulate public perception. In a recent incident \cite{bloombergmisinformation}, there was an instance of misinformation wherein a fabricated image of an explosion near the Pentagon, generated using AI technology, was disseminated through social media platforms. This dissemination of false information had notable repercussions, even influencing the US stock market, which experienced a brief dip as a result.


Motivated by the need to address the evolving threat landscape induced by GenAI, we propose leveraging the Cyber Kill Chain (CKC) as a foundational model for cyber defense. To the best of our knowledge, this paper represents the first in-depth investigation of cyber threats stemming from the adversarial use of GenAI within the context of CKC. The CKC framework originally developed by Lockheed Martin \cite{hutchins2011intelligence}, provides a comprehensive framework for understanding and disrupting the lifecycle of a cyberattack. By applying the CKC to GenAI-based attacks, we can identify and counteract each stage of the attack process, from initial reconnaissance to final exfiltration. This approach offers effective means of countering the evolving tactics employed by adversaries utilizing GenAI, ensuring a proactive defense against these emerging threats. 

In addition to countering the use of GenAI in cyberattacks, we also recognize the presence of defense-aware adversaries who adapt their strategies and techniques to circumvent traditional defensive measures. Numerous studies have shed light on the diverse strategies employed by defense-aware threat actors, encompassing evasion, polymorphism, deception, misinformation, etc. (see Section \ref{strategy} for details). These strategies, along with their variants, can be used in various permutations across several stages of CKC. 
  Research conducted by Khan et al. \cite{khan2021offensive}
demonstrates how threat actors can utilize state-of-the-art
language models to launch highly targeted spear-phishing campaigns during the reconnaissance phase, increasing the likelihood of successful compromise. Moreover, the work of Das and colleagues \cite{das2019automated} investigates the weaponization of GenAI generated text content, showcasing how malicious actors can exploit generative models to craft convincing social engineering messages, effectively bypassing traditional security defenses during the delivery phase. Using evasion strategies, threat actors can easily deceive existing defense systems. For example, AlEroud et al. \cite{aleroud2020bypassing} demonstrated the URLs generated as adversarial phishing examples in an inference integrity attack successfully bypassing the target classifier. Similarly, researchers have achieved success manipulating the Command and Control (C2) channel on social media platforms using generative models, highlighted in the works of\cite{rigaki2018bringing}.  

Furthermore, this paper proposes a taxonomy of defensive strategies that can be employed by organizations to enhance the resilience of cyber defense systems, effectively safeguard their critical assets, and protect their digital ecosystem from emerging threats induced by the offensive use of GenAI.

The major contributions of the paper are as follows:
\begin{itemize}
    \item Identification of the impact of offensive use of generative models on the cyber threat landscape, specifically in each phase of the Cyber Kill Chain (CKC) framework.
    \item Exploration of threat actors' strategies to shed light on how these actors employ sophisticated strategies to craft an attack.
    \item Taxonomizing defensive strategies for detecting and mitigating threats originating from generative models throughout the kill chain.
    \item Providing a foundation for understanding the emerging threats posed by the offensive use of generative models.
\end{itemize}

The remainder of the paper discusses the impact of GenAI in cybersecurity. In Section \ref{generative}, we provide an overview of the GenAI landscape including foundational concepts, taxonomy, and applications. Section \ref{strategy} discusses sophisticated attack strategies adopted by threat actors looking to maximize the success of an attack using GenAI. In Section \ref{kill_chain_risk}, we provide a comprehensive analysis of the risk introduced by the misuse of GenAI techniques in each phase of the CKC framework. Section \ref{defstrategy} recommends defense strategies that should be employed and embraced by organizations to mitigate exposure to adversarial attacks induced by GenAI. Finally, Section \ref{conclusion} reviews the major challenges posed by GenAI-based cyberattacks and discusses improvements to defensive strategies using emerging technologies. 

\section{Overview of Generative AI}\label{generative}

In this paper, we provide a high-level overview of the Generative AI (GenAI). For a deeper understanding, we direct our readers to refer to the NIPS 2016 Tutorial on Generative Adversarial Networks by Goodfellow et al \cite{goodfellow2016nips}. Furthermore, we recommend reviewing the article by Harshvardhan et al. \cite{ harshvardhan2020comprehensive} for comprehensive analysis, algorithms, architecture, and implementation of generative models.

The term GenAI is used to describe the use of unsupervised and semi-supervised machine learning algorithms to generate synthetic data from existing data such as text, audio and video files, images, or code, based on certain inputs. It is achieved through the utilization of machine learning model referred to as a \emph{generative model}, which is trained on a large dataset of examples and can produce new instances that are comparable to the ones it was trained on. For instance, given training data $\tilde p_{data}(x)$, the goal of generative model is to learn $\tilde p_{model}(x)$ and generate new samples from same distribution that approximates $\tilde p_{data}(x)$ as closely as possible. 

\begin{figure*}[ht]
    \centering
    \includegraphics[scale=.78]{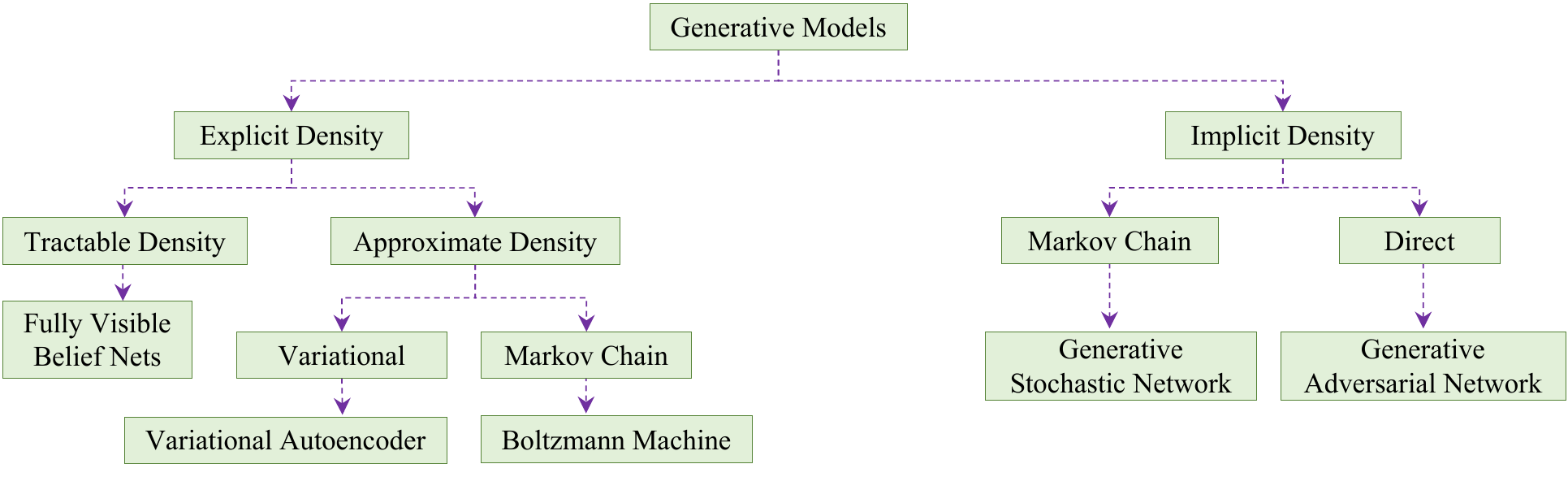}
    \caption{ Adapted from the work of Goodfellow et al. \cite{goodfellow2016nips}, a taxonomy of GenAI based on the tractability of their density distributions. The left branch represents GenAI, which utilizes explicit density estimation techniques as its foundation, whereas the right branch represents GenAI, which leverages implicit density estimation techniques.}
        \label{fig: Generative_models}
        \end{figure*}

One method for learning a close approximation of $\tilde p_{data}(x)$ involves defining an explicitly parameterized function $\tilde p_{model}(x; \theta )$ and iteratively searching for values of the parameters that maximize the similarity between $\tilde p_{data}(x)$ and $\tilde p_{model}(x)$. Most of the generative models are based on the principle of maximum likelihood estimation while minimizing the Kullback-Leibler divergence between $\tilde p_{data}(x)$ and $\tilde p_{model}(x)$. The idea is to choose the parameters for the model that maximize the likelihood of the training data.

A taxonomy of deep generative models is illustrated in Fig. \ref{fig: Generative_models}. The left branch of the taxonomy represents models that define explicit density functions, whereas the right branch represents models that define implicit density functions. Explicit density is further divided into two categories such as tractable density and approximate density. Fully Visible Belief Nets (FVBN) \cite{frey1995does} and Nonlinear Independent Component Analysis (NICA) are example of models based on tractable density functions. These models provides explicit parametric specification of the distribution of an observed random variable $x$ by defining a log-likelihood function $log q _\theta (x)$ with parameters $\theta$. However, with the emergence of Deep Learning (DL), the focus of study has shifted to learning models with relatively complex functional forms, because when DL is used for synthesis, the corresponding density function may be computationally intractable \cite{goodfellow2020generative}. 

To circumvent some of the limitations imposed by the design requirements of models with tractable density functions, other models have been developed that use deterministic approximations such as variational methods and those that use stochastic approximations such as Monte Carlo methods. One of the most popular variational deep generative models is Variational Autoencoder (VAE) \cite{kingma2013auto}. Another approximation method is Markov Chain Approximation. A Markov chain is a
process for generating samples by repeatedly drawing a sample $\boldsymbol{x}^{\prime} \sim q\left(\boldsymbol{x}^{\prime} \mid \boldsymbol{x}\right)$. By repeatedly updating $x$ according to the transition operator $q$, Markov chain methods can sometimes guarantee that $x$ will eventually converge to a sample from $p_{model}(x)$. However, the convergence with this methods is slow and computational cost is high. 

Implicit models, on the other hand, as illustrated on the right side of taxonomy Fig \ref{fig: Generative_models} eliminate the need to build a tractable density function. Instead, these models offer a way to train the model without directly interacting with $p_{model}$, usually by taking samples from it. One such popular method is Generative Adversarial Networks (GAN). Unlike other density models, GANs do not require numerous approximations or Markov chains and have the ability to generate a sample in a single step. 

\subsection*{GenAI \& Application Areas} 
Large language models, such Generative Pre-trained Transformer (GPT-3 \cite{devlin2018bert} and GPT-4) and Bidirectional Encoder Representations from Transformers (BERT) \cite{devlin2018bert}, are state-of-the-art transformer-based text generator models.
GANs and VAEs can also be used for text generation task but these models are not computationally efficient due to their underlying architecture compared to transformer models. Models for text-to-image generation include, DALL-E 2 
\cite{ramesh2021zero}, Imagen \cite{kingma2021variational}, and Midjourney \cite{midjourney}. While Phenaki \cite{villegas2022phenaki}, DreamFusion \cite{poole2022dreamfusion}, and Codex \cite{chen2021evaluating} are examples of models that can generate videos, 3D object, and code from text. Several works have also been proposed to generate audio using text as an input, for example AudioML \cite{borsos2022audiolm}, and Wishper \cite{radford2022robust}. 

In the following section, we will outline the attack strategy employed by an attacker. A comprehensive awareness and knowledge of the attacker's tactics will pave the way for the creation of efficient defensive measures.

\section{Attack Strategies leveraging GenAI} 
\label{strategy} 
Cyberattacks are constantly evolving, infiltrating all protection measures and stealthily exfiltrating sensitive data. 
Understanding how such attacks work assists in developing attack-aware defensive systems. However, the rapid advancement in technology, in particular, the emergence GenAI and its misuse (offensive use), has made attack vectors even more sophisticated, stealthy, and able to bypass defense systems relatively easily. Hence, it is hard to formulate a concrete plan to mitigate cyberattacks without understanding the attack strategies that threat actors employ and embrace. In this paper, we identify five such attack strategies: \emph{evasion, adaptation and automation, polymorphism,  deception}, and \emph{misinformation}. 

A clear picture, comprehension, and awareness of the abovementioned strategies can effectively help defenders build robust defensive systems. {However, defenders should be aware that these are some of the main strategies. There can be variants, combinations, and permutations of these strategies. Table \ref{table:attack_strategies_ckc_stage} summarizes how a combination of one or more attack strategies can be used in several phases of the Cyber Kill Chain (CKC). For example, an evasion attack strategy can be used to gather information about a network during reconnaissance, hide information using steganography or other obfuscation techniques to deliver a malicious payload, or install exploit rootkits and malware.

In addition, the defender needs to be aware of any variations on these attack strategies, for instance, a variant of evasion technique is the "\emph{Living-of-the-Land}" (LOTL) technique, which leverages processes and legitimate tools (PowerShell, Windows Management Instrumentation (WMI), Mimikatz, etc.) that are already present on systems to avoid detection, perform network reconnaissance, allow lateral movement, and maintain persistence}\cite{barr2021survivalism}. POSHSPY \cite{poshspy} {is an example of a LOTL-based attack employed by the Russian state-sponsored APT (Advanced Persistent Threat) group, APT29}.


\subsection{Evasion}
\label{evasion}

\begin{boxEnv}
    \textit{Evasion} is commonly used to describe the process of conducting an attack while evading or bypassing detection methods.
\end{boxEnv} For example, in a ransomware attack like WannaCry \cite{wannacry}, the attackers utilized an exploit called EternalBlue, which targeted a vulnerability in Microsoft Windows systems and utilized evasion techniques, including encryption and obfuscation, to evade detection and spread rapidly. Furthermore, evasion can be used to bypass firewalls, intrusion detection and prevention systems (IPS, IDS), and malware analysis. 

An evasion attack involves the creation of malicious inputs, commonly referred to as adversarial examples using GenAI, that cause the detection system to misclassify or make inaccurate predictions. 
Adversaries, for example, can utilize GANs to generate adversarial malicious traffic to attack intrusion detection systems, bypass them, and propagate throughout the network \cite{lin2022idsgan}. Similarly, they can also create malicious URLs that can easily bypass URL and web detectors undetected \cite{aleroud2020bypassing, apruzzese2022spacephish}. One of the evasion strategies in  the malware domain is the \emph{obfuscation technique}, a method of concealing the attack payload used by malware to circumvent static analysis methods and conventional anti-malware solutions. Adversaries can utilize GenAI to further optimize the obfuscation technique. For example, they can generate creative dead code insertion techniques, subroutine reordering (changing a code's subroutine to evade detection), register reassignment (changing a register to a newer generation while retaining program code and behavior), etc. Therefore, malware that has been obfuscated using generative models might be challenging to detect \cite{hu2023generating}.

\subsection{Adaptation and Automation} 
\label{adaptation} 
\begin{boxEnv}
    \emph{Adaptation} refers to the ability of a system or entity to adjust or modify itself in response to changes in its environment or circumstances. \emph{Automation}, on the other hand, refers to the use of technology to perform tasks or processes with minimal human intervention.
\end{boxEnv}
Examples of attacks involving adaptation and automation strategies include NotPetya \cite{petya}, a destructive malware designed with an automated propagation mechanism to infect new systems, while CrashOverride \cite{crashoverride}, a malware framework that targeted Industrial Control Systems (ICS) used in electric power grids, was designed to automate the various stages of Cyber Kill Chain (CKC).

Threat actors continuously augment their attacking skills through a combination of learning, collaboration, and the adoption of emerging technologies. In the past, threat actors relied on manual techniques and limited automation to launch attacks. For instance, manually composing phishing emails, manually building malware variants, manually researching possible targets, etc. While these approaches were partially effective, they were time-consuming and required a substantial amount of work to scale. At present, threat actors may adopt state-of-the-art Large Language Models (LLM) like GPT-3 \cite{brown2020language} and GPT-4 that can generate human-like language to generate personalized and highly convincing phishing messages that closely mimic legitimate communications and use them for reconnaissance. 

Such adaptation reduces the labor-intensive aspects of attacks, allowing threat actors to concentrate on more strategic planning and enabling them to scale their operations and launch attacks more rapidly.

\subsection{Polymorphism} 
\label{polymorphism}

\begin{boxEnv}
    \emph{Polymorphism} refers to the ability of an object or data element to take on different forms or types. In the context of network security, it refers to a strategy of self-mutation, or change in the form of attack vectors.
\end{boxEnv} 

An example of this strategy, according to Darkreading \cite{darkreading}, includes India's SideWinder Advanced Persistent Threat (APT) group targeting Pakistani government officials and individuals, using polymorphism techniques to bypass traditional signature-based antivirus detection and deliver a next-stage payload.  Examples of malware that utilize this strategy include Ursnif (or Gozi) \cite{ursnif} and AAEH (or Beebone) \cite{aaeh}, among others.

The attacks involving polymorphism mutate malware signatures in such a way that an anti-malware detection system fails to recognize the new signature variation, thereby infiltrating the target network or systems. Currently, malware can evolve into thousands of polymorphic versions. The defense against polymorphic malware comes in two flavors: static analysis and dynamic analysis \cite{alosefer2012analysing}. However, static analysis could be challenging due to the dynamic nature of malware and the size of the code bases that must be analyzed for polymorphic behavior.

Attackers can potentially misuse the ability of GenAI to generate synthetic data to craft polymorphic attacks. Chauhan et al. \cite{chauhan2020polymorphic} employed GANs to demonstrate that deep learning techniques were not sufficient for detecting new attack profiles. The same authors, in their later work, showcased that attackers can launch polymorphic Distributed Denial of Service (DDoS) attacks using a generative model such as WCGAN \cite{chauhan2021polymorphic}.

\label{skill}
\subsection{Deception}
\begin{boxEnv}
    \textit{Deception} is an interaction between two parties, a deceiver and a target, in which the deceiver successfully causes the target to accept as true a specific incorrect version of reality, with the intent of causing the target to act in a way that benefits the deceiver \cite{rowe2007deception}.
\end{boxEnv}
One example of trojan malware that employs deception strategies to evade detection is the Emotet trojan , which spreads via phishing email attachments and links that, once clicked, launch the payload. According to the MITRE ATT\&CK framework \cite{emotet}, Emotet evades detection by means of obfuscation and software packing.

GenAI, specifically LLMs, offer powerful capabilities for generating realistic and deceptive content, enabling attackers to manipulate, deceive, and exploit their targets. Offensive applications of LLMs include deceptive content generation for phishing emails and social engineering messages. Attackers, for example, can generate content that closely mimics the \emph{writing style, tone}, and \emph{vocabulary} of certain individuals or organizations, including relevant details and references, making it difficult for the recipient to identify the email as malicious. 

Threat actors can also use LLMs for \emph{impersonation} and \emph{identity theft}. For example, they leverage readily available tools such as ChatGPT \cite{chatgpt}, Brad \cite{gbrad} or Bing Chat \cite{bingchat} to generate deceptive messages, legal documents, or official-looking notices that imitate the communication style of a trusted individual or organization and use them to trick people into obtaining sensitive information or gaining unauthorized access to network systems.

\begin{table*}[ht]
{\renewcommand{\arraystretch}{1.20}%
\begin{tabular}{p{3.4cm}|p{3cm}|p{10.2cm}}
\hline
\rowcolor{cyan!20!}
Attack Strategy         & Cyber Kill Chain Stage                   & Select Example and Techniques  \\                                                                                          \hline

                                    \multirow{3}{*}{\makecell{Evasion\\}} & Reconnaissance & 

                                    \begin{minipage}[t]{\linewidth}
    \begin{itemize}[leftmargin=*]
        \item Network Scanning, OS fingerprinting. 
    \end{itemize} 
    \end{minipage}\\
                                    
 & Delivery & 

 \begin{minipage}[t]{\linewidth}
    \begin{itemize}[leftmargin=*]
        \item Steganography, or other obfuscation techniques.
    \end{itemize} 
    \end{minipage}\\
 
    & Installation & \begin{minipage}[t]{\linewidth}
    \begin{itemize}[leftmargin=*]
        \item Employ rootkit or stealth mechanism.
    \end{itemize} 
    \end{minipage}\\
    
    \hline

                  \multirow{3}{*}{\makecell{
Adoption and Automation\\}} & Delivery & 
 \begin{minipage}[t]{\linewidth}
    \begin{itemize}[leftmargin=*]
        \item Automated botnets to distribute malware or launch DDoS attack.
    \end{itemize} 
    \end{minipage}\\

    & Command and Control &  \begin{minipage}[t]{\linewidth}
    \begin{itemize}[leftmargin=*]
        \item Automated tools to manage and control compromised systems. 
    \end{itemize} 
    \end{minipage}\\
    
      & Exploitation & \begin{minipage}[t]{\linewidth}
    \begin{itemize}[leftmargin=*]
        \item Use of automated exploit tools to scan for vulnerabilities. 
    \end{itemize} 
    \end{minipage}\\
      
       \hline
                                    
 \multirow{4}{*}{\makecell{
Deception\\}} & Reconnaissance & 
 \begin{minipage}[t]{\linewidth}
    \begin{itemize}[leftmargin=*]
        \item Intelligence gathering about target (Social Engineering, Phishing). 
        \item Creation of fake personas to trick user.
    \end{itemize} 
    \end{minipage}
  \\ 
    & Exploitation & 
\begin{minipage}[t]{\linewidth}
    \begin{itemize}[leftmargin=*]
        \item Use of fake or manipulated data to exploit vulnerability in target systems (Injection of malicious code in legitimate website).
    \end{itemize} 
    \end{minipage}\\ 
    & Weaponization & \begin{minipage}[t]{\linewidth}
    \begin{itemize}[leftmargin=*]
        \item Malware disguised as legitimate files or applications, fileless attacks (e.g., Living-of-the-Land (LOTL)).
    \end{itemize} 
    \end{minipage} \\ 
    
      & Installation & 
      \begin{minipage}[t]{\linewidth}
    \begin{itemize}[leftmargin=*]
        \item Trick user via phishing message into installing payload.
    \end{itemize} 
    \end{minipage} \\ 
     \hline
              
\multirow{3}{*}{\makecell{
Polymorphism\\}} & Weaponization & \begin{minipage}[t]{\linewidth}
    \begin{itemize}[leftmargin=*]
        \item Creating of polymorphic malware.
    \end{itemize} 
    \end{minipage} \\

   & Delivery & \begin{minipage}[t]{\linewidth}
    \begin{itemize}[leftmargin=*]
        \item  Continuous change in delivery techniques using different file formats or encryption algorithms.
    \end{itemize} 
    \end{minipage} \\

    & Exploitation  & \begin{minipage}[t]{\linewidth}
    \begin{itemize}[leftmargin=*]
        \item   Polymorphic shellcode can continuously modify the exploit payload during an attack.
    \end{itemize} 
    \vspace{1mm}
  \end{minipage} \\ 
     \hline

                        \multirow{2}{*}{\makecell{
Misinformation\\}} & Reconnaissance & 
\begin{minipage}[t]{\linewidth}
    \begin{itemize}[leftmargin=*]
        \item  Spread false information to mislead and divert attention from real attack vectors.
    \end{itemize} 
    \end{minipage} \\
    & Action on Objectives & \begin{minipage}[t]{\linewidth}
    \begin{itemize}[leftmargin=*]
        \item  Create false report to cause panic or confusion. 
    \end{itemize} 
    \end{minipage} \\   \hline
                                    
                                    
\end{tabular}}
\\
\\
\caption{An overview of how threat actors can utilize attack strategies along the different stages of the Cyber Kill Chain (CKC) with some select examples and techniques in the attack process.}

\label{table:attack_strategies_ckc_stage}

\vspace{-4mm}

\end{table*}

\label{deception}
\subsection{Misinformation} 
\label{misinformation}
\begin{boxEnv}
    \emph{Misinformation} is false or inaccurate information that is deliberately created and is intentionally or unintentionally propagated \cite{wu2019misinformation}. It is an attack on our cognitive being. False information that is deliberately created and intentionally propagated to mislead is called \emph{disinformation}.
\end{boxEnv}
 Threat actors such as nation-state players with geopolitical ambitions, extremist groups, or economically motivated companies that exploit information use this strategy to create societal unrest, enhance polarization, and, in certain instances, affect the outcome of elections.
 The proliferation of social networks such as Twitter \cite{twitter}, Facebook \cite{facebook}, and Reddit \cite{reddit}, coupled with easy internet access, has created a fertile environment for the spread of misinformation. These platforms have facilitated the easy publication of content without proper editing, fact-checking, or accountability \cite{kumar2014detecting}. While cyberattacks focus on targeting computer networks and end systems, deliberate misinformation and half-truths can be employed to manipulate individuals' perspectives or divert unsuspecting users from accessing accurate information \cite{forbes2002web}.

 Furthermore, cutting-edge GenAI technology provides adversaries with the means to produce highly realistic fake content encompassing images, audio, videos, and text known as ``deepfakes''. These synthetic media capabilities are made feasible through the application of GANs, AutoEncoders and other GenAI models. Threat actors can exploit these technique to create realistic artificial faces to populate numerous bot accounts and use them to disseminate of misinformation \cite{ranade2021generating, khurana2019preventing, bertino2021ai,moroney2021case}.  Some popular deepfake tools available on the market for face swap include Deepswap \cite{deepswap}; for lip-syncing, Wombo \cite{woombo}; and for bot creation, Instagram Deep Bot \cite{instadfbot}.

{In the following section, we will explore how threat actors can utilize these strategies in conjunction with GenAI as a powerful offensive tool to launch cyberattacks. Specifically, our research will focus on investigating the implications of GenAI on the CKC, which serves as a framework that formalizes the study of cyber defense. By examining the perspective of an threat actor equipped with GenAI within the context of the CKC, this research intends to contribute to a deeper understanding of the challenges posed by advanced adversaries and aid in the development of effective security measures against GenAI-enabled cyberattacks.}

\section{Cybersecurity Risk Areas Impacted by GenAI} \label{kill_chain_risk}

Sun Tzu \cite{tzu2020art}, in his book The Art of War, once said, ``If you know the enemy and know yourself, you need not fear the result of a hundred battles.'' This quote encapsulates the fundamental importance of understanding both the attacker and oneself in achieving success in warfare.


\begin{figure}[h]
    \centering
    \includegraphics[scale=.70]{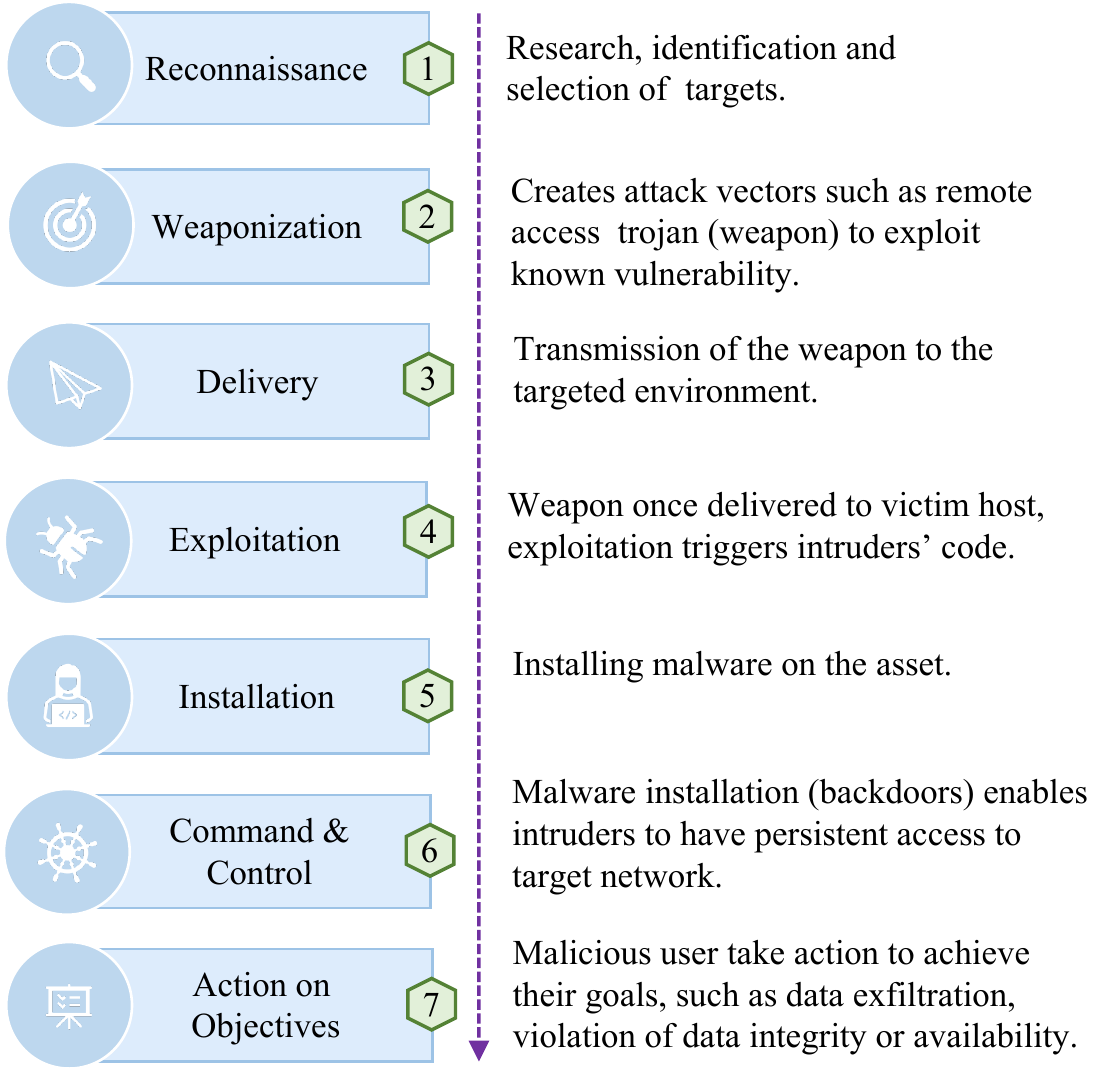}
    \caption{Adapted from Lockheed Martin \cite{hutchins2011intelligence}, Cyber Kill Chain (CKC) stages illustrate the sequential progression of a cyberattack from reconnaissance to action on objectives. Each stage represents a crucial step in the threat actors’ strategies and methodologies, providing insights for defensive countermeasures. }
        \label{fig: cyber_kill_chain}
\end{figure} 
In the realm of cybersecurity, this principle holds true, where risks abound. The ever-increasing risks and threats in the digital landscape necessitate a comprehensive approach to defense. To address the problem of evolving risks, the Cyber Kill Chain (CKC) \cite{hutchins2011intelligence} concept was developed. Inspired by Lockheed Martin's military model, CKC is an intelligence-driven framework for cybersecurity defense. It provides a systematic approach to identifying, preparing for, engaging with, and neutralizing cyber threats.

Drawing upon the analogy of Sun Tzu's quote, we can see a clear parallel between threat actors, defenders, and the CKC. The primary goal of CKC is to enhance cyber defense by thoroughly understanding the Tactics, Techniques, and Procedures (TTPs) employed by threat actors (knowing the enemy) while simultaneously identifying vulnerabilities and weaknesses within one's own systems (knowing yourself). Armed with this knowledge, defenders can develop effective strategies, implement robust countermeasures, and proactively prevent and mitigate the risk of cyberattacks.

\begin{figure*}[h]
    \centering
    \includegraphics[scale=.58]{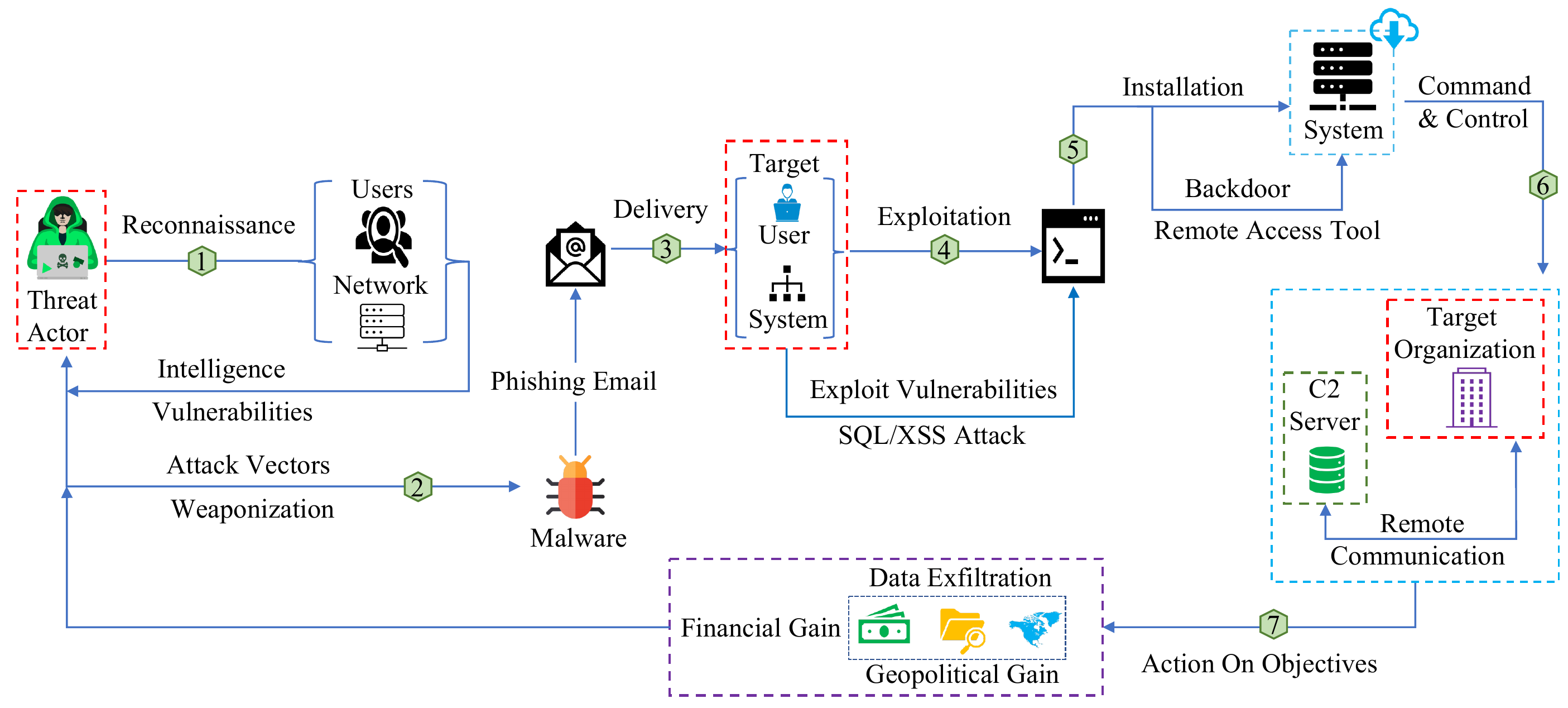}
    \caption{Tracing an attack utilizing the Cyber Kill Chain (CKC) attack framework. In the first stage, the threat actor gathers intelligence through \emph{reconnaissance}. The second stage is \emph{weaponization}, where threat actors create attack vectors (e.g., malware) based on identified vulnerabilities. The third stage is about \emph{delivering} a malicious payload in the target environment (usually with phishing and social engineering). A malicious payload is triggered in the fourth stage to \emph{exploit} vulnerabilities. In the fifth stage, backdoors are \emph{installed} to maintain undetected persistence. The sixth stage is about establishing \emph{command \& control} in the target environment for remote communication. Threat actors \emph{achieve their goals}, such as financial gain, data exfiltration, geopolitical gain, etc., after a successful attack in the final stage.}
        \label{fig: ckc_exploded}
\end{figure*}

The proposed CKC model, as indicated in Fig \ref{fig: cyber_kill_chain}, includes  seven distinct chained stages: \emph{reconnaissance, weaponization, delivery, exploitation, installation, command and control (C2)}, \& \emph{action on objectives}. Fig \ref{fig: ckc_exploded} illustrates the sequential stages of a cyberattack in high level. It begins with reconnaissance, followed by weaponization, where attack vectors like malware are created. The third stage involves delivering the payload through methods like phishing. Exploiting vulnerabilities triggers the payload in stage four. The fifth stage establishes undetected persistence with backdoors. The sixth stage establishes command \& control, enabling remote communication. The final stage sees the threat actors achieve their goals, such as financial gain or data exfiltration.

 {Our objective is to thoroughly examine how threat actors can exploit GenAI at each stage to carry out stealthy and sophisticated attacks. In the subsequent subsection, we will provide a detailed analysis of each stage, emphasizing the potential security implications and the use of GenAI from an offensive standpoint.}

\subsection{Reconnaissance} 
\label{recon}

\textit{Reconnaissance} is the first phase of a cyberattack in which threat actors study potential targets to find vulnerabilities. The efficacy of an attack improves as more intelligence is obtained on a target. Reconnaissance can be active or passive, depending on how the intelligence is acquired. Passive reconnaissance, sometimes called footprinting, involves target surveillance with a minimum amount of direct interactions. Using available Open-Source Intelligence (OSINT) tools, adversaries can identify and select their target, profile their system and social network, and validate their target.
Once attackers have gathered sufficient knowledge about the target through publicly available resources, they transition to \emph{active reconnaissance}. Active reconnaissance, sometimes called scanning, involves direct interactions with the target to gain additional insight.

Adept adversaries can leverage GenAI to improve both passive and active reconnaissance strategies. As noted in Section \ref{strategy}, GenAI augments traditional information retrieval techniques and facilitates the creation of sophisticated attacks. The success of ChatGPT has fueled an AI arms race that continues to improve the state-of-the-art in Large Language Models (LLMs). Major search engines are integrating AI chatbots (e.g., ChatGPT, Bing Chat and Bard) to assist users on queries, productivity, and content creation. In order to upstage competitors, LLMs are being pushed for early release which leads to gaps in safeguarding mechanisms due to reduced timelines for verification and validation (V\&V) procedures. Astute threat actors can use carefully-designed prompts to circumnavigate safeguards on LLMs. 

AI chatbots can be used to quickly generate specialized summaries on potential targets including organizational structure, personal details on users/employees, and network information. AI chatbots can also generate boilerplate code for scanning network traffic with little to no prior experience. The ability to simulate what-if scenarios by engaging (role-playing) with AI chatbots is another added benefit to adversaries. Through role-playing, game trees can be constructed to counter target defenses for different scenarios at different stages.

    \subsection{Weaponization}     \label{weapon}
    Once the vulnerabilities in the target system are identified during the reconnaissance step, the attackers will transition into the second stage, referred to as \emph{weaponization}. In this phase, the objective is to develop an attack that exploits the vulnerability discovered in the previous phase, where the choice of weapon depends on gathered information. Metasploit, Exploit-DB, Veil Framework, Aircrack, Burpsuit, etc. are some of the examples of weapons used in this phase that assist in formulating a deliverable payload, such as malwares, Remote Access Trojans (RATs) or backdoors coupled with exploits. 
    

    Polymorphic malware obfuscated with generative models is an excellent example of a weaponized payload that threat actors may use in this phase to exploit vulnerabilities. Threat actor can leverage GenAI to construct polymorphic malware, which can generate malicious code variants with unique characteristics that can avoid detection by traditional signature-based detection systems. To learn the patterns and structures of malicious code, for example, a generative model can be trained on a dataset of malware samples. GenAI can then produce new variants by altering certain properties, such as code control flow structure or encryption algorithms (refer to Section \ref{polymorphism} and \ref{evasion} for more information on polymorphism and obsfucation strategies).

    \subsection{Delivery} 
    \label{delivery}
    
The attacker must find a way to deliver its payload to its intended targets. During this stage, threat actors consider all possible avenues to deliver the malicious payload into the target's network or device. In the body of literature, the most common malicious payload delivery approaches observed are via auxiliary means such as phishing email attachments, social engineering, USB sticks, and drive-by download. In some cases, attackers can also directly attack a target’s web services.

A spear-phishing attack includes sending an email with a malicious attachment (payload) to a target victim, typically by impersonating an internal employee. Currently, deep learning along with language models are being applied to launch advanced spear phishing attacks. One such example of utilizing Recurrent Neural Networks (RNN) and Natural Language Generation (NLG) to generate spear-phishing attacks is presented by authors in \cite{das2019automated}. The proposed system generates fake emails with varying malicious payloads (depending on the target). Moreover, it has also been found that attackers can leverage state-of-the-art generative language models such as Generative Pre-Trained Transformers (GPT-2, GPT-3 and GPT-4) to automate the process of email generation. The work by Khan et al. \cite{khan2021offensive} highlights the use of a state-of-the-art language model for spear-phishing activity.

Sophisticated attackers could use GenAI to deceive (\emph{deception strategy}) the target's defense system and deliver a malicious payload. Anderson et al. \cite{anderson2016deepdga} show that GANs can be used to generate an undetectable malware URL that bypasses a deep neural network (DNN) based detection system. At present, several URL classification algorithms have been proposed to detect malicious URLs; however, these defensive mechanisms are susceptible to \emph{evasion} attacks using GenAI. AlEroud et al.'s \cite{aleroud2020bypassing} work precisely demonstrates how existing URL classification can be circumvented using GANs. 
Similarly, a recent work by the authors of \cite{apruzzese2022spacephish} presents a realistic threat model describing evasion attacks against machine learning-enabled phishing web detectors using GAN classifiers. 

    \subsection{Exploitation} \label{explaoitation}  

The attack payloads or malicious programs delivered in the preceding stage are triggered in the exploitation stage, enabling threat actors to exploit the vulnerabilities and weaknesses of the target systems and/or network. To gain better control and a stronger footing in the target environment, threat actors employ various exploitation techniques to compromise systems, networks, and applications. The most commonly used technique involves identifying and exploiting software vulnerabilities. Ghaffarian et al. \cite{ghaffarian2017software} define a \emph{software vulnerability} as ``an instance of a flaw caused by a mistake in the design, development, or configuration of software such that it can be exploited to violate some explicit or implicit security policy.'' A threat actor may exploit software vulnerabilities to gain unauthorized access or execute arbitrary code by attacking weak endpoints in software using techniques such as buffer overflows, SQL injection, cross-site scripting (XSS), and remote code execution.

Threat actors can leverage GenAI to automatically create synthetic SQL injection scripts and use them to attack web frameworks. One example of creating a synthetic SQL attack sample by leveraging GANs can be found in the work of \cite{lu2022gan}. The authors combine genetic algorithms and GANs to generate synthetic samples of SQL injection attack statements that are more complex and richer injection codes. XSS is another category of injection attack in which malicious scripts are injected into otherwise safe, benign, and trusted websites. Similar to the creation of adversarial SQL injection attack samples, threat actors can generate sophisticated and stealthier XSS attack samples that can bypass detection system(s) utilizing GenAI. For example, Zhang et al. \cite{zhang2020adversarial} constructed a GAN and proposed generating adversarial samples of Cross-Site Scripting (XSS) attacks with the Monte Carlo Tree Search (MCTS) algorithm. Samples are generated following a set of bypassing rules such as hexadecimal encoding, decimal encoding, url encoding, insertion of invalid characters in the middle of tags, and case mixture. This capability of GAN can be leveraged offensively by a threat actor to generate and automate XSS attacks.

\subsection{Installation} \label{installation} 

After successfully exploiting the vulnerability, threat actors, in this stage, attempt to gain access to additional nodes (i.e., spread the malware in the network) and install remote administration tools, such as Remote Access Trojans (RAT) or backdoors, in order to maintain their presence in the target environment \cite{hutchins2011intelligence} and exfiltrate valuable data. RAT refers to software, a piece of code, or a model that executes on the target’s system and provides remote, hidden, and undetected access to the threat actor, whereas exploits utilize RAT to maliciously exploit the vulnerabilities of a system or software to infiltrate or initiate an attack.

In the context of crafting backdoor attacks, also known as trojaning attacks, using generative learning techniques, the attacker teaches the model to recognize an unusual pattern that triggers a specific action, for example, classifying a sample as safe \cite{ mirsky2022threat}. One method for injecting a backdoor involves adding a Trojan trigger to a subset of the training sample and modifying the associated labels to correspond to the target class.  Pixels can serve as triggers in the spatial domain (images, videos), whereas words or a phrase can serve in the sequential domain \cite{azizi2021t}. GenAI can be used to create such triggers. For instance, Gu et al. \cite{gu2017badnets} presented the BadNets attack, in which the Trojan is injected by tainting the training dataset. In this framework, the attacker stamps a trigger pattern, i.e., a pixel, on an arbitrary subset of images in the training data. The attackers then mislabel the modified samples with the desired target label; consequently, whenever the trigger pattern is present, the learning classifier is trained to misclassify. In a similar fashion, the authors of \cite{liu2017trojaning} devised a trojaning technique that creates triggers and ensures that specific neurons are activated in order to inject a backdoor. 

Besides, GenAI also enable the creation of stealthy and evasive installation methods. For instance, GANs can generate steganographic techniques that hide malicious code within innocuous-looking files (images, text etc.), making detection and analysis more challenging for defensive systems. A novel technique called SteganoGAN for hiding arbitrary binary data in images using GANs that is capable of evading detection by steganalysis tools is proposed by Zhang et al. \cite{zhang2019steganogan}.

    \subsection{Command and Control (C2)} 
    \label{c2}

After an attacker has successfully compromised a system, they typically look to establish Command and Control (C2) in infrastructure systems, endpoints, etc., which allows them to remotely communicate with the compromised system and issue commands via C2 channels. If the attackers successfully execute the previous steps, their access to the system is persistent even if the compromised system is restarted or a patch is installed. The infected device can either immediately begin executing missions or wait for further instructions from its C2 server.

Attackers can utilize GenAI techniques to evade detection during the (C2) phase by creating models that mimic the network behavior of legitimate applications, camouflaging the traffic as normal. To illustrate this, Rigaki et al. \cite{rigaki2018bringing} presented a method employing GAN that modifies the behavioral patterns of a specific malware in the network to mimic Facebook chat traffic in order to conceal C2 communication. Results show that the proposed method can successfully modify the traffic of malware to make it undetectable. Other methods that hide (\emph{evasion strategy}) malware in neural networks to deliver malicious pay load and enable C2 communication channels are StegoNet \cite{liu2020stegonet} and EvilModel \cite{wang2022evilmodel}. 

DeepC2 \cite{wang2022deepc2} on the other hand, use deep learning techniques to launch a covert C2 scenario on online social networks such as Twitter. In order to create C2 links, Domain Generation Algorithms (DGAs) are utilized by a large number of malware families. Anderson et al, \cite{anderson2016deepdga} leverage the concept of GANs to construct a deep learning based hard-to-detect artificial malware domains DGA that can even bypass a deep learning-based detector. 
One of the techniques that is currently in practice to defend against C2 attacks is deep SSL packet inspection, a network packet inspection system that examines packet payloads to identify traffic flows (TCP/IP connections) in a C2 channel. However, detection of both encrypted and non-encrypted malware command and control traffic based on TCP/IP is vulnerable to evasion attacks using deep generative models. For example, generative models can be used to create realistic network traffic such as ICMP pings, DNS queries, and HTTP web requests and can be used to evade C2 detectors. Examples of use of GenAI to bypass detectors in the C2 channel are in \cite{novo2020flow, rigaki2018bringing}. 

    \subsection{Action on Objectives}
      \label{action on obj}

    The action on objectives is the last stage of the kill chain, wherein the attackers execute action to achieve their objective. The action is usually dictated by the motivation of the attacker. Understanding the type of attacker (a single person, an insider, a nation-state, etc.) is critical to understanding their motivation. Attackers may be motivated by several factors, including financial gain, political gain, nation-state interests, malicious insiders, or simply wanting to move laterally to go after a more important system in the network. Financial motivation is evident with attacks involving ransomware, which are designed to encrypt the target’s end system, deny custodians to access their data and systems, and demand ransom payments in exchange for releasing the system or encrypted data. Usually, these attacks are carried out by successful delivery of malware using sophisticated phishing strategies (malicious email attachments), drive-by downloads, malicious URLs, etc.

    A zero-trust defensive security model is designed with the assumption that every user account, device, or endpoint will fall victim to action on the objective stage of the cyber kill chain by removing the idea of trust and treating all traffic to the network as untrusted until proven otherwise (usually through a rigorous authentication process). Essentially, throughout this phase, defenders are actively monitoring user and device activity to identify potential threats. In contrast, adversaries could possibly model user behavior using GenAI. For example, they can first generate user behavior and later use evasion and deception strategies to bypass the zero-trust security model. An instance of abnormal user behavior generation using DCGAN in a zero-trust network is presented in the work of \cite{qu2022abnormal}. 

    Lateral movement is another important aspect in this phase of the CKC, where the adversary's end goal is to move laterally inside the network, perform reconnaissance about the deeper network, and launch an attack. One way of moving laterally without being detected by defensive systems is via dynamic IP mutation. In such a situation, adversaries can leverage GenAI techniques to mutate IP addresses dynamically. For instance, it is possible to produce mutant IP addresses in an unpredictable manner without compromising network performance \cite{al2013random}.

    Another objective of adversaries in this phase is to exfiltrate target data. Usually, adversaries use malware to compromise systems or exfiltrate data. GenAI can be used to obfuscate malware and hence bypass detection. An example of the use of a GenAI to evade malware detection is MalGAN \cite{hu2023generating}. Works like MalGAN showcase how adversaries can use state-of-the-art generative models to avoid detection, which consequently paves their way to achieve their goal of exfiltrating the data.

    Next, we explain defensive strategies. 

\section{Defense Strategies to Mitigate GenAI Risks} \label{defstrategy}

The relentless surge in cyberattacks, orchestrated by ever-evolving threat actors, has escalated both in frequency and complexity. As a result, conventional defensive strategies are proving woefully inadequate in the face of these persistent and highly skilled attackers, particularly when they exploit cutting-edge attack techniques, notably harnessing the power of GenAI. Reactive measures and commercial security solutions, such as patching end systems and software to prevent zero-day exploits and blocking bad IP addresses and domain names, are the primary defense mechanisms employed by most organizations. However, these measures are insufficient against advanced attacks that leverage deep learning techniques, specifically, GenAI and sophisticated attack strategies (refer to  Section \ref{strategy}). Therefore, there is a critical need to develop proactive defense strategies that can detect and respond to these evolving threats.

\begin{figure}[h]
    \centering
    \includegraphics[scale=.85]{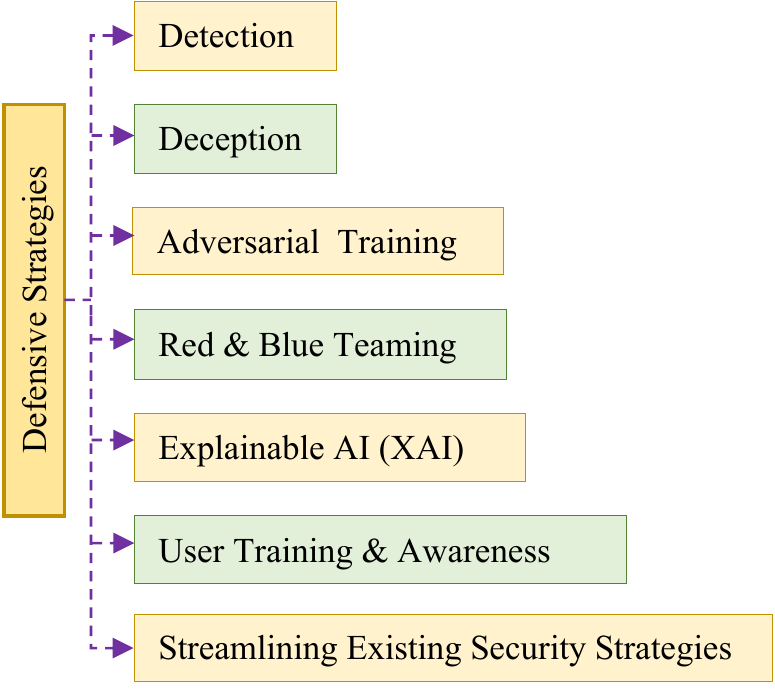}
    \caption{A taxonomy of proposed defensive strategies to mitigate GenAI risk, including \emph{detection, deception, adversarial training, red and blue teaming, explainable AI}\emph{user awareness training}, and \emph{streamlining existing security strategies}. The combination of these strategies aims to enhance the resilience of systems and networks against cyberattacks.}
        \label{fig: genai_defense}
               \end{figure}

To develop an effective defensive framework, it is crucial to understand the attacker's strategy, their knowledge, and their goal. In the preceding Section \ref{kill_chain_risk}, we explored the capabilities of GenAI from the perspective of a sophisticated threat actor. Drawing on this knowledge, 
we present a taxonomy of defensive strategies, as depicted in Fig \ref{fig: genai_defense}, encompassing \emph{detection, deception, adversarial training, red and blue teaming, explainable ai}, \emph{user awareness training} and \emph{streamlining existing security strategies} aimed at countering the risks introduced by GenAI throughout the Cyber Kill Chain (CKC). 



In the following sections, we will provide detailed explanations for each of these categories.





\label{defensive_strategies}

\subsection{Detection}

GenAI is increasingly being used as a detection technique to mitigate organizational exposure to cyberattacks. Systems developed with these models can learn the underlying distribution of normal network or system behavior and detect deviations from that distribution that may indicate an attack. 

Several works can be found in the body of literature that focus on unsupervised \emph{anomaly detection} by leveraging generative models such as GANs. For example, the work in TadGAN \cite{geiger2020tadgan}, TanoGAN \cite{bashar2020tanogan}, and Mad-GAN \cite{li2019mad} represents anomaly detection in time series data. 

Building on the success of anomaly detection tasks using GenAI, researchers have also explored their application in other cybersecurity domains, such as phishing detection.
\emph{Phishing attacks} are a common method used by attackers to obtain sensitive information from individuals and organizations. These attacks can be used in \emph{reconnaissance} and \emph{delivery} phase of CKC. 
Shirazi et al. \cite{shirazi2020improved} demonstrated the generative capability of adversarial auto-encoders in generating synthesized phishing samples when augmented with real-world phishing data provided additional robustness to phishing detection classifiers. Another work showcases the effectiveness of the combination of GAN (LeakGAN) and large language models (BERT \cite{devlin2018bert}) in phishing email detection \cite{qachfar2022leveraging}. 


In addition to combating phishing attacks, the application of GenAI extends to the field of \emph{malware detection}, where these models have proven instrumental in identifying polymorphic and metamorphic malware that constantly adapt to evade conventional detection methods. 
Using the Deep Belief Network (DBN), Yuan et al. \cite{yuan2016droiddetector} presented a system for Android malware detection referred to as \emph{DroidDetector} that can consequently distinguish whether a file has malicious behavior or not. GANs have also been used to detect malware on Android devices, such as in work of \cite{amin2022android}.

Moreover, as the field of cybersecurity continues to evolve, researchers have extended the utilization of GenAI beyond anomaly, phishing and malware detection to address the emerging challenge of deepfakes. \emph{Deepfakes} are synthetic videos, images, or texts that are created using generative models and can be used to manipulate or deceive individuals and organizations. To tackle this issue, researchers have developed detection systems tailored to different types of deepfakes. Fagni et al. \cite{fagni2021tweepfake} proposed a novel text-based social media message (Twitter) deepfake detection system. They demonstrated that the RoBERTa \cite{liu2019roberta} detector outperforms both traditional ML models (e.g., bag-of-words) and neural network models (e.g., CNN) in successfully distinguishing machine-generated tweets from human-written tweets. Another study on the detection of fake social media texts utilizing GPT-2 was conducted by Adelani et al. \cite{adelani2020generating} on Amazon product reviews. Intel most recently launched FakeCatcher \cite{fakecatcher}, a technology that can detect deepfake videos with a 96\% accuracy in real-time.




\subsection{Deception}

Defensive deception, for example, the deployment of decoy assets and honeypots—a network of seemingly vulnerable machines designed to lure threat actors—is a widely studied area in the security domain and is an established technique often used to gather Cyber Threat Intelligence (CTI), such as indicators and signatures of cyberattacks across various network environments \cite{kotheimer2016using,mittal2016cybertwitter,pingle2019relext,piplai2020creating}. This strategic technique has also been explored to set up a trap to distract and mislead attackers from real systems and thwart threat actors' (both insiders' and outsiders') efforts to identify and steal sensitive information on systems or networks \cite{voris2015fox}.

Researchers are currently investigating generative models to create sophisticated honeypots, or in some cases, honeyusers. For instance, Lukas et al. \cite{lukas2021deep} propose the creation of fake user accounts as honey tokens (honeyusers) on Active Directory Server using Variational Auto Encoder (VAE) to capture malicious access attempts. While some researchers are working to optimize and transform the log reports regenerated by honeypots such as Cowrie into question-and-answer problems using language models such as GPT-2 \cite{setianto2021gpt}.

\subsection{Adversarial Training}

In the context of cyberattacks, \emph{adversarial training} is used to defend against adversarial examples, which are small perturbations to input data that is specifically designed to deceive or mislead a machine learning model. 

This strategy has been shown to be effective in defending against various types of cyberattacks, including malware and phishing attacks. Several GAN-based studies have proposed methods to improve the evasion detection of adversarial malware. Rafiq et al. \cite{rafiq2022investigation} demonstrated the vulnerability of Android malware classifiers in adversarial settings. They showcased an adversarial training scheme to combat evasion attacks on ML-based Android malware classifiers with 99.46\% detection accuracy. In a similar vein, Taheri et al. \cite{taheri2020can} employed GANs to develop a defense against evasion attacks, employing five distinct evasion attack models on Android malware classifiers. 

Other defensive strategies include \emph{defensive distillation}, a technique used to improve the robustness of a machine learning model against adversarial attacks by creating a distilled model. Papernot et al \cite{papernot2016distillation}. were the first to demonstrate that defensive distillation may withstand adversarial attacks with minor perturbations in white-box settings.

\subsection{Red and Blue Teaming}

Red Teaming is a strategy used in military and commercial security operations to discover networked system vulnerabilities or exploitable gaps in operational concepts, with the main goal of decreasing surprises, improving them, and ensuring their robustness \cite{choo2007automated}. Blue Teaming, on the other hand, aims to protect networked systems and infrastructure from both simulated and real-world attacks while preserving the security posture of an organization.

The significance of red and blue teaming resides in their combined efforts to identify vulnerabilities, replicate real-world attack scenarios, and develop defensive mechanisms. Red teams can emulate a phishing attack scenario using language models or create the evasive malware using state-of-the-art generative models and launch it in a controlled environment to test the resilience of their networked systems. One example of maximizing the success of penetration testing with effective reconnaissance using ChatGPT is presented by the authors in \cite{temara2023maximizing}. First-hand exposure to such sophisticated attacks, even in a simulated environment, will empower the blue team with experience and prepare them to develop proactive countermeasures to effectively defend against such threats should they arise in the real world. Collaboration and feedback between red and blue teams helps firms better understand their security gaps, develop more effective incident response strategies, and strengthen their defenses over time.

\begin{figure*}[h]
    \centering
    \includegraphics[scale=.68]{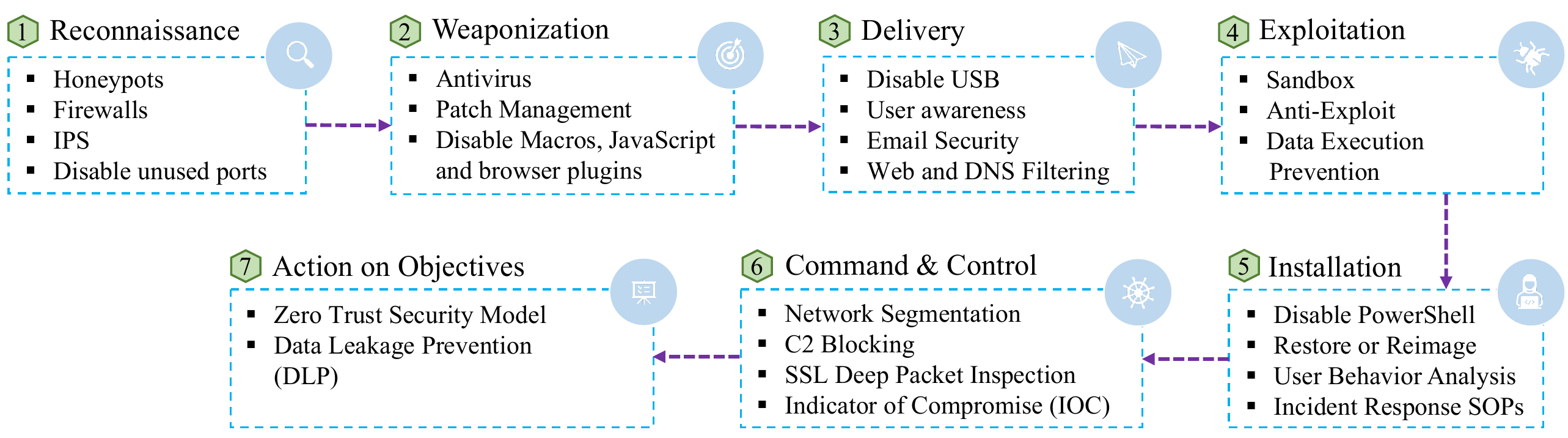}
    \caption{Recommended defensive tools and techniques for each phase of the Cyber Kill Chain (CKC) framework that should be adopted by organization in conjunction with GenAI-enabled security strategies for comprehensive cyber defense.}
        \label{fig: cybersecurity_defense_tools}
\end{figure*}

\subsection{Explainable AI (XAI)}
Explainable Artificial Intelligence (XAI) refers to the research and implementation of Artificial Intelligence (AI) models and systems that can provide human-comprehensible explanations for the output generated by ``black box'' AI models. Researchers are increasingly using XAI as a cybersecurity defense strategy to safeguard networks and endpoints. For example, in our previous works, we investigated XAI techniques in the intrusion detection domain \cite{neupane2022explainable, ables2022creating, kirby2023pruning} and explained the anomalies using feature importance in vehicular digital twins \cite{neupane2023twinexplainer}. 

XAI provides interpretability and transparency in AI models, and its integration can aid organizations in detecting and identifying deceptive content and adversarial examples, assessing model vulnerabilities, and monitoring emerging threats. Greco et al. \cite{ greco2023explaining} presented an XAI-based solution to classify phishing emails (deceptive content) and alert users to the risk by explaining the reasons behind the attacks. In another study by Morcos et al. \cite{morcos2022surrogate}, the authors utilized SHAP methods to enhance transparency and interpretability in malware detection (exfiltration detection). Fidel et al. \cite{fidel2020explainability}, in their work, presented a novel method to detect adversarial input using SHAP methods. Furthermore, in the case of Advanced Persistent Threat (APT) detection, XAI techniques were utilized by the authors in \cite{li2021explainable}.


\subsection{User Training and Awareness}

The importance of security awareness and training resides in the fact that it equips individuals with the knowledge and skills required to identify and mitigate the associated with offensive use generative models. As the sophistication of generative models increases, they have the potential to generate content that is both realistic and deceptive, which can be used to launch an attack using social engineering or phishing, create malware or fake identities, or even manipulate sensor data in cyber-physical systems. By offering comprehensive security awareness and training programs, organizations can educate their employees, and system operators about these common attack vectors generated by the misuse of generative models. This in turn, enables them to recognize the signs of manipulated or fraudulent content, develop a security-conscious mindset and become first line of defense. 

Overall, organizations can fortify their cyber defenses, ensure the confidentiality, integrity, availability, and reliability of their networked systems, and create a workforce that is vigilant, proactive, and capable of effectively mitigating cyber threats by continuously reinforcing security awareness and conducting regular training sessions.

\subsection{Streamlining Existing Security Strategies}




In the preceding subsections, we emphasized the substantial impact of GenAI in bolstering cybersecurity defenses, encompassing vulnerability identification, attack detection, and response mechanisms. Nevertheless, it is vital to acknowledge that GenAI alone cannot serve as a comprehensive solution for cyber defense, and it should be integrated seamlessly with other existing security strategies. We strongly recommend adopting the defensive tools and techniques that we have identified, as illustrated in Fig. \ref{fig: cybersecurity_defense_tools}, to effectively mitigate the likelihood of falling victim to cyber risks at each stage of the Cyber Kill Chain (CKC).



Mitigation of the \emph{reconnaissance} phase entails the implementation of several defensive techniques. First, organizations should thoroughly review their website description and other online platforms to ensure that sensitive information about the company's infrastructure, systems, or employees is not publicly disclosed. This includes sanitizing the website description by removing or minimizing specific details that could be potentially exploited by attackers. Additionally, closely monitoring network traffic and analyzing system logs are vital steps to detect any suspicious activity. Disabling any unused ports and limiting the amount of information made available about the system and its users are also effective measures to prevent successful social engineering attacks.


In order to mitigate the likelihood of exposure during the \emph{weaponization} phase of a cyberattack, defenders can utilize several tools and techniques, such as anti-malware software to detect known threats, patching known vulnerabilities to prevent exploitation, and disabling potentially risky elements such as office macros, JavaScript, and browser plugins. Furthermore, limiting user privileges can prevent the installation and exploitation of malware.

Organizations can use a variety of tools and methods to make it less likely that they will be exposed during the \emph{delivery} phase of a cyberattack. These include using email and web filtering to stop malicious URL attachments, disabling USB devices, and training and educating users on how to spot phishing emails and suspicious websites.

Defensive tools and techniques to curb the risk of \emph{exploitation} include data execution prevention through the use of least privilege access control mechanisms, anti-exploit measures, and sandboxing. A careful consideration of these tools and measures aids organizations in preventing the exploitation of vulnerabilities and the execution of malicious code within their systems.

Organizations can reduce the likelihood that a threat actor will \emph{install} a malicious payload in a network by taking defensive measures such as disabling PowerShell in the Windows environment, analyzing user behavior in the network to identify anomalies, establishing incident response procedures, and having the ability to restore or reimage compromised systems. 

Network segmentation can effectively prevent threat actors from establishing \emph{C2} in the inner network by restricting their access to other network nodes. Additionally, the installation and implementation of network traffic monitoring tools (for example, SSL deep packet inspection) can be an effective way to detect and prevent C2 traffic. Lastly, implementing data leakage prevention systems to detect and prevent the exfiltration of sensitive and confidential data, along with a zero-trust security model and employing encryption to protect data in transit, can assist in curbing the effects of a cyberattack in the last phase of the Cyber Kill Chain (CKC).

\section{Conclusion \& Discussion} 
\label{conclusion}

As GenAI becomes more ubiquitous, we in this paper shed light on its impact on the cyber threat landscape. The ability to generate highly realistic and deceptive content such as text, image, video when used offensively can threaten the security and the trustworthiness of systems that use these technologies. We have explored how the malicious use of GenAI can exploit vulnerabilities and craft sophisticated cyberattacks, further emphasizing the importance of robust defense mechanisms.

To counter evolving cyber threats, we have proposed a cyber defense strategy built upon the Cyber Kill Chain (CKC) framework. By understanding and disrupting each stage of the attack lifecycle, organizations can effectively mitigate the potential risks posed by GenAI-induced cyber threats. Moreover, we have emphasized the importance of comprehending defense-aware adversaries and their strategies in the context of CKC. Threat actors can employ sophisticated strategies such as evasion, adaptation, automation, polymorphism, deception, and misinformation to bypass traditional defense mechanisms. Understanding these strategies, their combinations, permutations, and variants across several stages of CKC enables the development of more resilient defense systems.


Drawing upon our findings, we have recommended a range of defensive strategies and tools to aid defenders in mitigating threats. We explored defensive strategies from two standpoints: First, we developed a taxonomy of defensive strategies enabled by GenAI, encompassing detection, deception, adversarial training, red and blue teaming, explainable AI, user training and awareness programs, and streamlining existing security strategies. Second, we emphasized the need to use available tools and techniques to improve organizational security posture. It is important to recognize that defense systems enabled by GenAI alone are not sufficient to combat the evolving cyber threat landscape; they must be used in conjunction with available tools to realize a robust defense framework.


Most of the reactive and preventive security solutions that are at our disposal today focus on enhancing security policies, security monitoring, and installing robust access control mechanisms to combat cyber threats, but these solutions are inadequate to address the increasing speed, scale, and sophistication of automated cyberattacks. For example, the sheer volume of automated cyberattacks we see and hear almost in daily basis, has now surpassed the ability of humans to manually analyze them.

Therefore, there is now a critical need to develop automated defense solutions within organizations to address the problem of this evolving threat landscape. One such solution to this problem is the concept of Automated Cyber Defense (ACD) systems—a system that can self-discover, prove, and correct software vulnerabilities in real-time without human intervention.

Emerging ML technologies such as Reinforcement Learning (RL) that can learn from its own experiences by exploring its environment can be utilized to create defense agent systems with self-learning capabilities \cite{cardellini2022irs,piplai2020using}. For example, Applebaum et al. \cite{applebaum2022bridging} and Cardellini et al. \cite{cardellini2022irs} leveraged RL techniques to train an agent that is able to autonomously defend a system, minimizing self-damage from responses that use noisy sensor data. Nguyen et al. \cite{nguyen2021deep}, on the other hand,  surveyed several deep RL methods for autonomous defense strategies, such as autonomous RL-based IDS and multi-agent RL-based game theory approaches, to obtain optimal policies in different attacking scenarios. 

By staying vigilant, investing in advanced detection technologies, and fostering a culture of cybersecurity awareness, organizations can bolster their resilience and effectively safeguard against the growing threats introduced by the offensive use of GenAI.

\section*{Acknowledgment}
This work was supported by PATENT Lab (Predictive Analytics and TEchnology iNTegration Laboratory) at the Department of Computer Science and Engineering, Mississippi State University. 
\ifCLASSOPTIONcaptionsoff
  \newpage
\fi

\bibliographystyle{unsrt}
\bibliography{refs}

\end{document}